# MUSIC TEMPO ESTIMATION ON SOLO INSTRUMENTAL PERFORMANCE

*Zhanhong He, [1] Roberto Togneri, [1] Xiangyu Zhang [2]*

[1] University of Western Australia, Perth, Australia
[2] University of New South Wales, Sydney, Australia
zhanh.he.uw@gmail.com

## ABSTRACT

Recently, automatic music transcription has made it possible to convert musical audio into accurate MIDI. However, the resulting MIDI lacks music notations such as tempo, which hinders its conversion into sheet music. In this paper, we investigate state-of-the-art tempo estimation techniques and evaluate their performance on solo instrumental music. These include temporal convolutional network (TCN) and recurrent neural network (RNN) models that are pretrained on massive of mixed vocals and instrumental music, as well as TCN models trained specifically with solo instrumental performances. Through evaluations on drum, guitar, and classical piano datasets, our TCN models with the new training scheme achieved the best performance. Our newly trained TCN model increases the Acc1 metric by 38.6% for guitar tempo estimation, compared to the pretrained TCN model with an Acc1 of 61.1%. Although our trained TCN model is twice as accurate as the pretrained TCN model in estimating classical piano tempo, its Acc1 is only 50.9%. To improve the performance of deep learning models, we investigate their combinations with various post-processing methods. These post-processing techniques effectively enhance the performance of deep learning models when they struggle to estimate the tempo of specific instruments.

## 1. INTRODUCTION

Tempo is a fundamental attribute in modern music. It measures the speed of a piece of music in beats per minute (BPM). In digital sheet music, tempo determines the playback speed of notes, thereby determining the duration of audio. When using paper sheet music, humans cannot adhere to the required speed precisely, yet tempo still serves as a guide for performance speed. Tempo estimation is one of the earliest tasks in music information retrieval (MIR). Accurate tempo estimation is beneficial for various downstream tasks in MIR, with music genre classification and music recommendation being the most mentioned [1]. However, the role of tempo estimation in automatic music transcription (AMT) is less discussed.

The end goal of AMT is transcribing musical audio to sheet music [2]. Apart from musical notes, sheet music comprises a multitude of musical notations such as tempo, and key signature. However, current state-of-the-art AMT systems can only capture the musical notes [3]. As a compromise solution, these AMT systems output MIDI-like files. This is because MIDI only requires the notes and allows for other notations (e.g., tempo) to be omitted. Since tempo is essential for MIDI playback, it defaults to 120 BPM when not specified. This enables the MIDI file to be played but is not conducive to sheet music conversion.

The missing tempo in MIDI can be estimated by some music notation software, such as MuseScore [4]. However, due to the use of simple algorithms, significant deviations are frequently observed. While a similar function exists in pretty_midi [5], this empirical method published in 2001 [6] is outdated. In contrast, estimating tempo directly from audio has been the dominant approach since 2004 [7], which avoids the error accumulated by stepwise estimation.

In this paper, we investigate the feasibility of tempo estimation techniques in providing accurate tempo for music transcription. Previous research has focused on music composed of vocals and a mixture of different musical instruments. This is because tempo estimation commonly shares the dataset with genre classification [1], which requires a wide coverage of diverse music. Over a decade, representative tempo estimation methods have not focused on a single instrument. Therefore, the applicability of these methods to instrument recordings is questionable. We implement the state-of-the-art tempo estimation model, the multitask TCN proposed in [8], and re-train it with solo performance datasets of classical piano [9], drums [10], and acoustic guitar [11], respectively. To achieve optimal performance, we attempt several new combinations of deep learning models and post-processing probabilistic models, inspired from previous research [8].

## 2. RELATED WORKS

Tempo is derived from the period (i.e., time interval) between consecutive beats [12]. Therefore, in most cases, the positions of beats must be estimated prior to the tempo. The task of identifying beat positions in audio is known as beat tracking. As early as 1994 [13], Goto and Muraoka manually labelled onsets in the audio, and then used empirical algorithms to identify the beats. Relying on onset signal strength (OSS) functions for beat tracking was popular in the early 2000s [6][7]. OSS methods based on Fourier analysis, autocorrelation and filters are first applied in detecting onsets, then post-processing methods are used to identify which onsets are corresponding to the beats.

Subsequently, machine learning (ML) and deep learning (DL) models replaced the roles of OSS functions. ML/DL models estimate the approximate positions of beats (e.g., a beat located at 0.75s 10%, 0.76s 80%, 0.77s 5%, …), called the beat activation. Then, post-processing methods are applied to obtain accurate beat positions. ML models, particularly the Gaussian mixture model (GMM), were widely used between 2010 and 2015 [1]. DL models were also applied to beat tracking during the same period, with the recurrent neural network (RNN) being first introduced by S. Bock in 2011 [14]. Since then, many DL models have been developed, but those using the temporal convolutional network (TCN) have demonstrated the best performance so far [8][15][16][17].

There are two types of post-processing methods to deduce the tempo after beat tracking. One is to use probabilistic models such as the hidden Markov model (HMM) [18], dynamic Bayesian network (DBN) [19], or conditional random field (CRF) [20] to convert the beat activation into beats, and then estimate the tempo based on the beats. Another is to skip the post-processing of beats, estimating the tempo from the beat activation by DBN [19], CRF [20], or other probabilistic models. Instead, some DL models [8][15][21] can directly estimate the tempo from audio, bypassing beat tracking and achieved the state-of-the-art results.

To the best of our knowledge, prior research in the last decade has rarely focused on musical instruments. A rare example is [22], where beat tracking models were studied on datasets containing or not containing drums. Both [22] and other previous works are trained on mixed music datasets such as HJDB [23] and Gtzan [24]. It can be anticipated that they would perform poorly on instrumental performances, as demonstrated by [22] achieving only 30% beat F1 score on predicting classical piano [9]. We evaluate state-of-the-art tempo estimation models with post-processing methods on solo instrumental datasets, to address these limitations.

## 3. EXPERIMENTAL DESIGN

### 3.1. Datasets

We independently train and evaluate tempo estimation models using three solo instrumental datasets. The datasets used in this study are all high-quality audio sampled at 44.1kHz, and their details are as follows.

*3.1.1. Aligned Scores and Performances (ASAP)*

The ASAP dataset [9] contains 519 performances extracted from the MAESTRO dataset [25]. MAESTRO is collected from the international Piano-e-Competition, featuring solo piano performances by famous pianists. ASAP supplements the MAESTRO with beat and downbeat annotations. Since the ASAP does not provide tempos, we infer tempos from the beats and use these as ground truth for our task. To facilitate future comparisons, we follow the predefined train-test split of MAESTRO for the ASAP. As the duration of audios ranged from 2 to 30 minutes, we segment them into roughly 30-second clips. We use the aligned MIDI to inform the segmentation, ensuring that long silences are skipped, and segments are not made during a sustained note. The resulting training set comprises 4826 clips, while the test set includes 1301 clips, with a total duration of 44.3 hours.

*3.1.2. Groove MIDI Dataset (GMD)*

The GMD dataset [10] contains 1150 drum solos performed by intermediate to professional drummers. These drummers play strictly to a metronome, resulting in precise beats/downbeats and tempo annotations for the dataset. Similarly, we adopt the predefined train-test split and segment the audio files exceeding 30 seconds. Audio segments and original recordings that are less than 5 seconds are discarded. As the result, the training set comprises 1074 segments, the test set comprises 306 segments, and the total duration is 13.2 hours.

*3.1.3. GuitarSet*

The GuitarSet dataset [11] contains 360 high-quality recordings of solo guitar performances with detailed annotations, including beats, downbeats, and tempo. The dataset features recordings from 6 professional guitarists, with all recordings fixed to 30 seconds. Despite having a duration of only 3 hours, GuitarSet is well-organised and requires no pre-processing. We conduct the 4:1 train-test split with a random seed of 1234, resulting in a training set of 288 clips and a test set of 72 clips.

### 3.2. Data Augmentation

As there is a relatively small amount of data in the GMD and GuitarSet, we apply the data augmentation proposed in [8] to ensure a good DL model performance. Prior to being fed into models, the audio signals are transformed into spectrograms via a short-time Fourier transform (STFT). Changing the frames per second (FPS) of the STFT can alter the overlap of the Fourier window and produce more diverse spectrograms for the same audio input. In addition to the standard FPS of 100, we augment the dataset by adding FPS values of 95 and 105, providing three different spectrograms from the same audio, tripling the amount of training data.

### 3.3. Selected Models

In this experiment, we select the TCN model proposed by Bock *et al.* [8] as the baseline due to its modular structure [1] and competitive performance [17]. We implement the TCN model following [8] and separately train it on the ASAP, GMD, and GuitarSet datasets targeting the model on specific instruments. To clarify actual performance, we compare (1) our implemented and newly trained TCN model, with (2) the original pretrained TCN model [8], and (3) the pretrained RNN model [14], referring to Figure 1.

Due to the multitask architecture, both the pretrained TCN and our trained TCN model can simultaneously output the beat activation and tempo activation. Tempo activation (i.e., tempo approximations and their respective likelihoods, see Figure 1) can be accurately determined by smoothing histogram bins and performing interpolation [1], directly detecting the tempo without the beats.

On the contrary, there are also various post-processing methods to estimate the tempo from the beat activation. One approach is to use probabilistic models to directly estimate the tempo from the beat activation. The models we selected for this approach include the auto-correlation function (ACF) [14], DBN [19], and comb filter [26]. Another approach is to first estimate the beats using probabilistic models, and then estimate the tempo from beats using a simple inference function. We select CRF [20], DBN [19], and comb filter [26] to estimate the beats from the beat activation function, with the tempo inference function from [1].

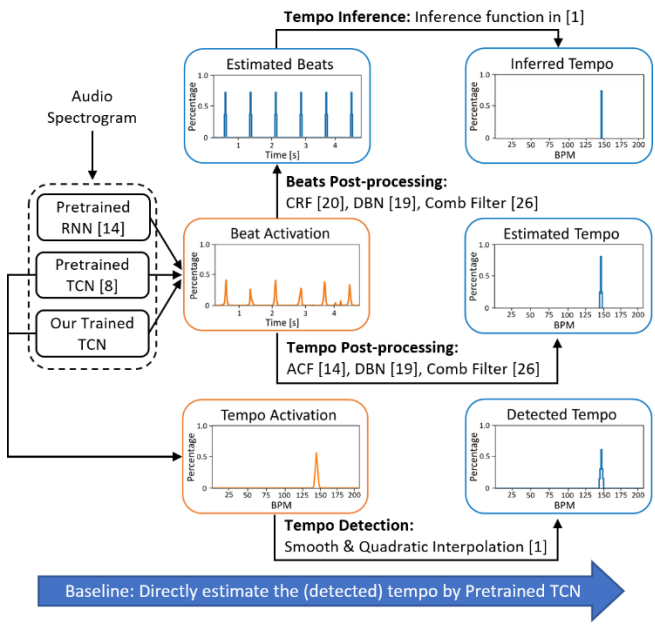

**Figure 1:** Three ways to obtain the tempo with a pretrained RNN/TCN and our trained TCN to (1) infer the tempo based on the beats, (2) estimate the tempo on the beat activation, and (3) detect the tempo based on the estimated tempo activation.

### 3.4. Network Training Scheme

We trained three separate TCN models, each using one of the ASAP, GMD, or GuitarSet datasets, to strengthen the performance on specific instruments. When implementing the TCN model, we strived to reproduce the architecture detailed in [8], as well as their training details. Given that the data varied in length from 5 to 30 seconds, we fed the TCN model with sequences of batch size 1 to avoid excessive padding, ensuring that the training data of different lengths could be accepted by the model. For model fitting, we utilize the combination of RAdam and Lookahead optimizers provided in TensorFlow-addons, as well as the cross-entropy loss designed for multiple estimation targets in [1]. Due to the chosen optimizers, the learning rate gradually decreases during the fitting process. We set the initial learning rate to 0.0015 and assign a clip gradient norm of 0.5. Different datasets require a different number of epochs to achieve a good fit. With early stopping, the ASAP and GMD dataset take about 150 epochs, while GuitarSet takes around 200 epochs. For each dataset, we repeat the training process four times and average the results to avoid overfitting or unexpected occurrences. The training was performed on the Nvidia RTX3080Ti 16G graphics card and 32G of RAM.

### 4. EVALUATION RESULTS

We evaluate models mentioned above to compare them using the metrics of Accuracy1 (Acc1) and Accuracy2 (Acc2). The Acc1 and Acc2 have been widely reported in the literature since being established as tempo evaluation metrics in 2004 [27]. Acc1 validates whether the estimated tempo is within +/- 4% tolerance of the annotated tempo. Building on Acc1, Acc2 allows for the reported tempo to be 1/3, 1/2, double, or triple the annotated tempo (i.e., tempo octave error in music terminology). We consider Acc1 as the strict standard to evaluate our models, while Acc2 serves as an auxiliary metric, as the tempo octave error is generally unacceptable in practical applications [27]. The evaluation is performed independently on the testing set of each dataset, and all results have been summarized in Table 1.

We first compare our trained TCN model with the baseline pretrained TCN model for directly detected tempo (underlined italicized values in Table 1). Except for the insignificant difference on the GMD dataset, our trained TCN model outperforms the pretrained TCN model on other reported datasets. Compared to the TCN baseline, our trained TCN model achieves over a 30% Acc1 improvement on the GuitarSet (99.7%), as well as double the Acc1 on the ASAP (50.9%) compared to the baseline TCN (61.1% in GuitarSet, 24.6% in ASAP). This indicates the necessity of using instrument-specific trained models for estimating tempo of instrument-only recording. One possible reason why our trained TCN does not outperform the baseline on the GMD is due to the difference in training data. The TCN baseline was trained on multiple datasets [8], most of which contained drums [22], whereas our TCN is solely trained on the GMD dataset. Drum features are prominent in mixed music [28], so training with mixed music or drum solos may lead to similar tempo estimation performance. This suggests that the training data amount is likely the determinate factor of drum tempo estimation.

| DL Model & Post-processing Method | | ASAP Acc1 | ASAP Acc2 | GMD Acc1 | GMD Acc2 | GuitarSet Acc1 | GuitarSet Acc2 |
|---|---|---|---|---|---|---|---|
| Pretrained RNN | Cannot Directly Det. Tempo | - | - | - | - | - | - |
| | CRF Beats Inf. Tempo | 0.423 | 0.696 | 0.552 | 0.974 | 0.500 | 0.889 |
| | DBN Beats Inf. Tempo | 0.332 | 0.661 | 0.631 | 0.987 | 0.653 | 0.889 |
| | Comb Filter Beats Inf. Tempo | **0.429** | 0.699 | 0.680 | 0.977 | 0.611 | 0.875 |
| | ACF Beat Act. Est. Tempo | 0.382 | 0.709 | **0.804** | 0.961 | **0.708** | 0.861 |
| | DBN Beat Act. Est. Tempo | 0.314 | 0.626 | 0.631 | 0.990 | 0.653 | 0.875 |
| | Comb Filter Beat Act. Est. Tempo | 0.410 | 0.710 | 0.683 | 0.980 | 0.667 | 0.903 |
| Pretrained TCN | Baseline - Directly Det. Tempo | _0.246_ | 0.671 | _0.879_ | 0.974 | _0.611_ | 0.917 |
| | CRF Beats Inf. Tempo | 0.391 | 0.694 | 0.670 | 0.971 | 0.611 | 0.917 |
| | DBN Beats Inf. Tempo | 0.304 | 0.655 | 0.693 | 0.987 | 0.625 | 0.889 |
| | Comb Filter Beats Inf. Tempo | **0.402** | 0.705 | 0.712 | 0.967 | 0.583 | 0.903 |
| | ACF Beat Act. Est. Tempo | 0.295 | 0.703 | 0.784 | 0.944 | 0.486 | 0.861 |
| | DBN Beat Act. Est. Tempo | 0.289 | 0.627 | 0.693 | 0.987 | **0.639** | 0.903 |
| | Comb Filter Beat Act. Est. Tempo | 0.365 | 0.689 | 0.735 | 0.977 | 0.542 | 0.889 |
| Our Trained TCN | Directly Det. Tempo | _0.509_ | 0.732 | _0.875_ | 0.933 | _0.997_ | 0.997 |
| | CRF Beats Inf. Tempo | 0.504 | 0.718 | 0.870 | 0.953 | 0.893 | 0.986 |
| | DBN Beats Inf. Tempo | 0.495 | 0.709 | **0.893** | 0.958 | 0.875 | 0.983 |
| | Comb Filter Beats Inf. Tempo | **0.529** | 0.729 | 0.869 | 0.951 | 0.900 | 0.972 |
| | ACF Beat Act. Est. Tempo | 0.515 | 0.745 | 0.822 | 0.948 | 0.844 | 0.976 |
| | DBN Beat Act. Est. Tempo | 0.473 | 0.674 | 0.886 | 0.952 | 0.872 | 0.986 |
| | Comb Filter Beat Act. Est. Tempo | 0.525 | 0.725 | 0.856 | 0.950 | 0.924 | 0.976 |

**Table 1:** Evaluation results for the Pretrained RNN, the Pretrained TCN, and Our Trained TCN. Evaluation is performed on the test sets of ASAP, GMD, and GuitarSet datasets respectively across all three methods presented in Section 3.3 (for detected, estimated and inferred tempo). The underline italicized Acc1 values are the TCN models directly detecting tempo, and the bold values are the best Acc1 scores achieved by the model across all processing methods.

Subsequently, we compare the performance of various post-processing methods applied on the DL models. Due to the complexity of classical piano music, our evaluated TCN models performed poorly on the ASAP dataset when directly detecting the tempo (i.e., both the TCN baseline 24.6% and our TCN 50.9% are quite low Acc1 scores). In such cases, applying the post-processing probabilistic models effectively enhances these DL models. As shown in the ASAP results, the combination of the pretrained TCN and comb filter improves the Acc1 by 11.8% (estimated tempo) and 15.6% (inferred tempo). Both post-processing approaches, inferring tempo from beats and estimating tempo from beat activation, are effective in improving the performance of the DL model. Inferring tempo from beats is validated on the pretrained TCN, increasing Acc1 from 24.6% (directly detected tempo) to 40.2% (inferred tempo via comb filter). Estimating tempo from the beat activation is validated on our trained TCN, increasing Acc1 from 50.9% (directly detected tempo) to 52.9% (estimated tempo via comb filter). We cannot conclusively determine which probabilistic model is most useful. Both DBN and comb filter are helpful for the pretrained TCN and our trained TCN, while ACF is more effective for the pretrained RNN, as shown by their performance on the three datasets.

Lastly, we have observed that there was little difference in the Acc2 among all evaluated models on the same dataset. The ambiguity of Acc2 makes it difficult to use for model comparison. Given that Acc1 was sufficiently accurate in this experiment, we are not using the Acc2 for comparing models.

## 5. CONCLUSION

In this paper, we examined the tempo estimation techniques with three solo instrumental datasets. Across these datasets, tempo estimation models trained for specific instruments significantly improve the performance, compared to training on massive mixed music. We suggest using models that have been trained specifically on each musical instrument, when supplementing the missed tempo of MIDI obtained from music transcription. In addition, we explored various post-processing methods to be applied with DL models. These post-processing models effectively complement the DL models, especially when estimating the tempo of classical piano music. Both post-processing approaches, estimating tempo from beat activation, and first estimating beats from beat activation and then inferring tempo, have performed well in different scenarios. We hope that future research on improving DL models will also consider incorporating post-processing models, as they can be simple yet effective ways to enhance tempo estimation.